\documentclass[10pt]{article}
\usepackage{latexsym}
\usepackage{amssymb}
\usepackage{amsfonts}
\usepackage{amsmath}
\usepackage{theorem}

\newtheorem{theorem}{Theorem}
\newtheorem{lemma}{Lemma}

\newtheorem{definition}{Definition}
\newtheorem{remark}{Remark}
\newtheorem{proposition}{Proposition}

\def\be{\begin{equation}}
\def\ee{\end{equation}}
\def\bc{\begin{center}}
\def\ec{\end{center}}

\begin{document}

\title{Stability properties and probability distributions of
  multi-overlaps in dilute spin glasses} \author{ Adriano Barra
  \footnote{King's College London, Department of Mathematics, Strand,
    London WC2R 2LS, United Kingdom, and Dipartimento di Fisica,
    Universit\`a di Roma ``La Sapienza'' Piazzale Aldo Moro 2, 00185
    Roma, Italy, {\tt<Adriano.Barra@roma1.infn.it>}} , Luca De Sanctis
  \footnote{ICTP, Strada Costiera 11, 34014 Trieste, Italy,
    {\tt<lde\_sanc@ictp.it>}}}

\maketitle

\begin{abstract}
We prove that the Aizenman-Contucci relations, well known for
fully connected spin glasses, hold in diluted spin glasses as
well. We also prove more general constraints in the same spirit
for multi-overlaps, systematically confirming and expanding
previous results. The strategy we employ makes no use of
self-averaging, and allows us to generate hierarchically  all such
relations within the framework of Random Multi-Overlap Structures.
The basic idea is to study, for these structures, the consequences
of the closely related concepts of stochastic stability,
quasi-stationarity under random shifts, factorization of the trial
free energy. The very simple technique allows us to prove also the
phase transition for the overlap: it remains strictly positive
(in average) below the critical temperature if a suitable
external field is first applied and then removed in the
thermodynamic limit. We also deduce, from a cavity approach, the
general form of the constraints on the distribution of
multi-overlaps found within Quasi-Stationary Random Multi-Overlap
Structures. 
\end{abstract}



\section{Introduction}

Dilute spin glasses are studied mostly for two reasons: their
finite connectivity makes them in a certain sense close to
finite-dimensional systems, while retaining a mean-field
character; and they are mathematically equivalent to some
important random optimization problems (such as X-OR-SAT and K-SAT
\cite{talabook}). The proper setting for the study of mean field
dilute spin glasses are the Random Multi-Overlap Structures
(RaMOSt), and the whole physics behavior of dilute spin glasses is
carried by the probability distribution of the multi-overlaps
\cite{lds1}, which play the same role as the 2-overlap does for
fully connected models. In the case of the latter, it is known
that the Ghirlanda-Guerra identities \cite{gg} allow for the
computation of the critical exponents governing the critical
behavior of the 2-overlap \cite{abds}, and guarantee that the
2-overlap is positive below the critical temperature
\cite{talabook}. The relations due to Aizenman and Contucci
\cite{ac}, on the other hand, imply \cite{abds} that the
expectation of the 2-overlap is strictly positive below the
critical temperature (due to a phase transition triggered by an
external field). Ghirlanda-Guerra identities are a consequence of
the self-averaging of the energy density, and extend to dilute
spin glasses, where one can also find more general relations for
multi-overlaps \cite{ldsf}. By contrast, Aizenman-Contucci (AC)
relations are a consequence of stochastic stability \cite{ac,bds},
but they also follow from a certain kind of self-averaging, as
shown by Franz et al. \cite{flt}, who extended the AC relations to
dilute spin glasses and multi-overlaps. In this paper we provide a
new proof of the AC relations for dilute spin glasses, and of
their generalized version for multi-overlaps. We emphasize that
stochastic stability (and similar concepts) are intimately related
to self-averaging properties. Moreover, both approaches can be
used with observable of various forms, not just the with the
energy. The joint use of the techniques developed in \cite{abds}
within the approach developed in \cite{lds1} is at the basis of
the present work. The latter is organized as follows. The next two
sections introduce the model and our notations, and illustrate the
cavity perspective which the RaMOSt approach relies on. In section
\ref{positivi} we show that simple symmetry arguments within the
cavity method lead to the proof of the phase transition of the
expectation of the overlap: below its critical temperature
the overlap remains strictly positive, if an external
(cavity) field is applied and then removed in the thermodynamic
limit. Section \ref{qs} is devoted to a proof that AC relations
hold in dilute glasses, along with relations in the same spirit
for multi-overlaps. We have already stressed that our proof is
radically different from the one hinted at in \cite{flt}. Section
\ref{barra} presents the form of the derivative with respect to a
perturbing parameter of the expectation of a generic function of
some replicas. The result makes it possible to develop
systematically the constraints on multi-overlaps, whose critical
behavior control can be here improved as compared to section
\ref{positivi} (although the critical exponents are not found
yet). 
The
straightforward but tedious and long calculations needed in some
expansions are reported in the appendices, preceded by concluding
remarks.


\section{Model and notation}

Consider $N$ points, indexed by latin letters $i,j,$ etc.,
with an Ising spin attached to each of them, so to have
spin configurations
$$
\sigma:\{1,\ldots ,N\}\ni i\to \sigma_{i}=\pm 1\ .
$$
Hence we may consider $\sigma\in\{-1,+1\}^{N}$.
Let $P_\zeta$ be a Poisson random variable of mean $\zeta$,
let $\{J_\nu\}$
be independent identically distributed copies of a random variable
$J$ with symmetric distribution. For the sake of simplicity we
will assume $J=\pm 1$, without loss of generality \cite{gt1}.
We want to consider randomly chosen points, we
therefore introduce $\{i_\nu\},\{j_\nu\}$ as
independent identically distributed random variables, with uniform
distribution over $1,\ldots,N$.
Assuming there is no external field,
the Hamiltonian of the Viana-Bray (VB) model for dilute mean
field spin glass is the following symmetric random variable
\begin{equation*}
H_N(\sigma, \alpha; \mathcal{J})=
-\sum_{\nu=1}^{P_{\alpha N}} J_\nu \sigma_{i_\nu}\sigma_{j_\nu}\ ,\
\alpha\in\mathbb{R}_+\ .
\end{equation*}
$\mathbb{E}$ will be the expectation
with respect to all the (quenched) variables, 
i.e. all the random variables
except the spins, collectively denoted by $\mathcal{J}$.
The non-negative parameter $\alpha$ is 
called {\sl degree of connectivity}.
The Gibbs measure $\omega$ is defined by
$$
\omega(\varphi)=\frac{1}{Z}\sum_{\sigma}
\exp(-\beta H(\sigma))\varphi(\sigma)
$$
for any observable $\varphi:\{-1,+1\}^{N}\to \mathbb{R}$,
and clearly
$$
Z_{N}(\beta)=\sum_{\sigma}\exp(-\beta H_{N}(\sigma))\ ,
$$
which is the well known partition function.
When dealing with more than one configuration, the product
Gibbs measure is denoted by $\Omega$, and various
configuration taken from the each product space are called
``replicas''. As already done above,
we will often omit the dependence on $\beta$ and on
the size of the system $N$ of various quantities.
In general, we will commit some slight
notational abuses to lighten the expressions when there
is no risk of confusion. The free
energy density $f_{N}$ is defined by
$$
-\beta f_{N}(\beta)=\frac1N\mathbb{E}\ln Z_{N}(\beta)\ .
$$

The whole physical behavior of the model is encoded by \cite{lds1}
the even multi-overlaps $q_{1\cdots 2n}$, which are functions of
several configurations $\sigma^{(1)},\sigma^{(2)},\ldots$ defined by
\begin{equation*}
q_{1\cdots 2n}=\frac1N \sum_{i=1}^N
\sigma_i^{(1)}\cdots\sigma_i^{(2n)}\ .
\end{equation*}


\section{Cavity approach and Random Multi-Overlap Structures}

The thermodynamic limit of the free energy density
exists if and only if the sequence of the 
increments (due to the addition
of a particle to the system) is convergent in the Ces\`aro sense
(indicated by a boldface {\bf C}):
$$
\lim_{M\to\infty}\frac1M\mathbb{E}\ln Z_{M}\equiv
\lim_{M\to\infty}\frac1M\sum_{n=0}^{M-1}
\mathbb{E}\ln\frac{Z_{n+1}}{Z_{n}}
\equiv \mathbf{C}\lim_{M\to\infty}\mathbb{E}\ln\frac{Z_{M+1}}{Z_{M}}\ .
$$
The idea at the basis of the cavity approach is in fact to measure
the effect on the free energy of the addition of one spin to the system
(see \cite{ass2} for a beautiful summary).
Let us denote the given $M$ spins by $\tau$, as we want to save
the symbol $\sigma$ for the added spin(s). Now, following \cite{lds1},
we can write, in distribution,
\begin{equation}\label{step}
-H_{M+1}(\tau, \sigma; \alpha)\sim
\sum_{\nu=1}^{P_{\alpha \frac{M^{2}}{M+1}}}
J_{\nu}\tau_{k_{\nu}}\tau_{l_{\nu}}
+\sum_{\nu=1}^{P_{\alpha \frac{2M}{M+1}}}
\tilde{J}_{\nu}\tau_{m_{\nu}}
\sigma_{i_{\nu}}\ ,
\end{equation}
where we have neglected a term which does not contribute when $M$ is
large \cite{lds1}, $\{\tilde{J}_{\nu}\}$ are independent copies of
$J$; $\{k_{\nu}\}$, $\{m_{\nu}\}$, and $\{l_{\nu}\}$ are independent
random variables all uniformly distributed over $\{1,\ldots, M\}$;
$\{i_{\nu}\}$ are independent random variables uniformly distributed
over the set $\{1,\ldots, N\equiv 1\}$, consisting of $\{1\}$ only. So
$\sigma_{i_{\nu}}\equiv \sigma_{1}$.  Notice that we can also write,
in distribution,
\begin{equation}\label{acca1}
H_{M+1}(\tau, \sigma; \alpha)\sim H_{M}(\tau;\alpha^{\prime})
+\tilde{h}_{\tau}\sigma_{1}
\end{equation}
where
$$\alpha^{\prime}=\alpha\frac{M}{M+1}\ ,\
\tilde{h}_{\tau}=-\sum_{\nu=1}^{P_{2
\alpha^{\prime}}}\tilde{J}_{\nu}\tau_{k_{\nu}}\ .
$$
Notice also that similarly
\begin{equation}\label{acca2}
H_{M}(\tau;\alpha)=H_{M}(\tau;\alpha^{\prime}, \mathcal{J})
+H_{M}(\tau;\alpha^{\prime}/M, \mathcal{\hat{J}})\ ,
\end{equation}
thanks to the additivity property of Poisson variables,
and the two Hamiltonians in the right hand side have
independent quenched random variables $\mathcal{J},\mathcal{\hat{J}}$.
Hence, if we call
$$
H_{M}(\tau;\alpha^{\prime}/M, \mathcal{\hat{J}})
=\hat{H}_{\tau}(\alpha^{\prime})=
-\sum_{\nu=1}^{P_{\alpha^{\prime}}}\hat{J}_{\nu}
\tau_{k_{\nu}}\tau_{l_{\nu}}\ ,
$$
then
$$
\mathbb{E}\ln\frac{Z_{M+1}}{Z_{M}}
=\mathbb{E}\ln\frac{\sum_{\tau,\sigma}
\xi_{\tau}\exp(-\beta \tilde{h}_{\tau}\sigma)}{\sum_{\tau}
\xi_{\tau}\exp(-\beta \hat{H}_{\tau})}\ ,
$$
with
$$
\xi_{\tau}=\exp(-\beta H_M(\tau;\alpha^{\prime}))\ .
$$
As elegantly explained in \cite{ass2}, this equation expresses
the incremental contribution to the free energy in terms of the
mean free energy of a particle (a spin) added to a reservoir
whose internal state is described by $(\tau, \xi_{\tau})$,
corrected by an inverse-fugacity term $\hat{H}$,
which encodes a connectivity shift.
The latter may be thought of as the free energy
of a ``place holder'': the {\em cavity} into which the
$(M + 1)$st particle is added.
One may note that the addition of a particle to the reservoir of
$M$ particles has an effect on the state of the reservoir.
For $M >> 1$, the value of the added
spin, $\sigma$, does not affect significantly the
field which would exist for the next increment in $M$.
Hence, for the next addition
of a particle we may continue to regard the state of the
reservoir as given by just the
configuration $\tau$. However, the weight of the
configuration (which is still to be normalized
to yield its probability) undergoes the change:
$$
\xi_{\tau}\to\xi_{\tau}e^{-\beta \tilde{h}_{\tau}\sigma}\ .
$$
This transformation is called {\sl cavity dynamics}.

When we add more particles to the system, they
do not interact, as there will just be copies of the
cavity fields $\tilde{h}^{i}$ acting 
paramagnetically on each added spin
(see \cite{lds1} for details).
Therefore if we add infinitely many
particles (to an already infinite reservoir), we can replace
the initial complicated model with a simpler (at least in principle)
paramagnet.
The reasoning just illustrated thus paves
the way to the proper concept to introduce
for the computation of the free energy \cite{lds1}.
\begin{definition}
  Given a probability space $\{\Omega,\mu(d\omega)\}$,
  a {\bf Random Multi-Overlap Structure}
  $\mathcal{R}$ is a triple
  $(\Sigma, \{\tilde{q}_{2n}\}, \xi)$ where
  \begin{itemize}
  \item $\Sigma$ is a discrete space;
  \item $\xi: \Sigma\rightarrow\mathbb{R}_+$
    is a system of random weights, such that $\sum_{\gamma\in\Sigma}
    \xi_\gamma\leq\infty$ $\mu$-almost surely;
  \item $\tilde{q}_{2n}:\Sigma^{2n}\rightarrow\mathbb{R},
    n\in\mathbb{N}$
    is a positive semi-definite \emph{Multi-Overlap Kernel}
    (equal to 1 on the diagonal of
    $\Sigma^{2n}$, so that by Schwartz inequality $|\tilde{q}|\leq 1$).
  \end{itemize}
\end{definition}
By looking at the properties of $\tilde{h}, \hat{H}$
in (\ref{acca1})-(\ref{acca2}), we know that when many particles
(say $N$) are added to the system, we
need \cite{lds1} in general $N+1$ random variables
$\{\tilde{h}^{i}_{\gamma}(\alpha; \tilde{J})\}_{i=1}^{N}$ and
$\hat{H}_{\gamma}(\alpha; \hat{J})$, $\gamma\in\Sigma$,
such that
\begin{eqnarray}
  \frac{d}{d\alpha}\mathbb{E}\ln \sum_{\gamma\in\Sigma}\xi_{\gamma}
  \exp(-\beta\tilde{h}_\gamma^{i})
  & = & 2\sum_{n>0}\frac{1}{2n}\tanh^{2n}(\beta)
  (1-\langle \tilde{q}_{2n}\rangle)\label{eta} \\
  \frac{d}{d\alpha}\mathbb{E}\ln \sum_{\gamma\in\Sigma}\xi_{\gamma}
  \exp(-\beta \hat{H}_\gamma)
  &=& \sum_{n>0}\frac{1}{2n}\tanh^{2n}(\beta)
  (1-\langle\tilde{q}^{2}_{2n}\rangle)\ .\label{kappa}
\end{eqnarray}
These are the fields to plug into the ``trial pressure''
($-\beta f_N(\beta)$)
\begin{equation}\label{gtfdilute}
  G_{N}(\mathcal{R})=\frac 1N\mathbb{E}\ln
  \frac{\sum_{\gamma, \sigma}\xi_{\gamma}
    \exp(-\beta\sum_{i=1}^{N}\tilde{h}^{i}_{\gamma}
    \sigma_{i})}{\sum_{\gamma}\xi_{\gamma}
    \exp(-\beta \hat{H}_{\gamma})}\ .
\end{equation}

The Boltzmann RaMOSt \cite{lds1} is the one we started from,
constructed by thinking of a reservoir of $M$ spins $\tau$
$$
\Sigma=\{-1,1\}^M\ni\tau\ ,\ \xi_\tau=\exp(-\beta H_M(\tau))\ ,\
\tilde{q}_{1\cdots 2n}=\frac1M\sum_{k=1}^M\tau^{(1)}_k
\cdots\tau^{(2n)}_k
$$
with
$$
\tilde{h}_\tau^{i}(\alpha)=\sum_{\nu=1}^{P_{2\alpha}}
\tilde{J}_{\nu}^{i}\tau_{k_{\nu}^{i}}\ , \
\hat{H}_\tau(\alpha N)=
-\sum_{\nu=1}^{P_{\alpha N}}
\hat{J}_\nu \tau_{k_\nu}\tau_{l_\nu}
$$
and all the $\hat{J}$'s are independent copies of $J$,
independent of any other copy.

The next theorem will not be used, but it justifies the whole
machinery described so far.
\begin{theorem}[Extended Variational Principle]
Infimizing for each $N$
separately the trial function $G_N(\mathcal{R})$ over the whole
RaMOSt space, the resulting sequence tends to the limiting pressure
$-\beta f$ of the VB model as $N$ tends to infinity:
\begin{equation*}
\label{evp-d}
-\beta f=\lim_{N\rightarrow\infty}
\inf_{\mathcal{R}}G_{N}(\mathcal{R})\ .
\end{equation*}
\end{theorem}

A RaMOSt $\mathcal{R}$ is said to be optimal if
$G(\mathcal{R})=-\beta f(\beta)\ \ \forall\ \beta$.
We will denote by $\Omega$ the measure
associated with the RaMOSt weights $\xi$ as well.

Is it possible to show \cite{lds1} that
optimal RaMOSt's enjoy the same
factorization property enjoyed by the Boltzmann
RaMOSt, described in the next
\begin{theorem}[Factorization of optimal RaMOSt's]
\label{lisboa-d}
In the whole region where the parameters are uniquely
defined, the following Ces{\`a}ro  limit
is linear in $N$ and $\bar{\alpha}$
\begin{equation*}\label{limrost}
\mathbf{C}\lim_{M}\mathbb{E}\ln\Omega_M
\{c_{1}\cdots c_{N}\exp[-\beta \hat{H}(\bar{\alpha})]\}
=N(-\beta f +\alpha A)+\bar{\alpha}A\ ,
\end{equation*}
where $c_{i}=2\cosh(\beta\tilde{h}^{i})$,
\begin{equation}\label{defa}
A=\sum_{n=1}^{\infty}\frac{1}{2n}
\tanh^{2n}(\beta)(1-\langle q_{2n}^{2}\rangle)\ ,
\end{equation}
and the averages in both sides of the equation
are assumed to taken by means of weights at connectivity $\alpha$.
\end{theorem}
As we extended the use of $\Omega$ from the Gibbs measure to any
RaMOSt measure, we are clearly extending to any RaMOSt the notation
$\mathbb{E}\Omega(\cdot)=\langle\cdot\rangle$ too.  We will use only
part of the previous theorem, or more precisely a modification of the
part involving the inverse fugacity only.  Notice that in the
definition (\ref{gtfdilute}) of the trial pressure $G$ the part with
the cavity fields and the part with the inverse fugacity are taken
already factorized (the inverse fugacity appears at the denominator).
If we therefore focus on the fugacity part only in the theorem above,
by setting all the cavity fields $\tilde{h}^{i}$ to zero, the property
described in Theorem \ref{lisboa-d} becomes what is often called
\textsl{stochastic stability} (of the measure $\Omega$ with respect to
the perturbation $\hat{H}$), which we will prove and exploit in
Section \ref{qs}.  It should be now clear, from the construction
described about equations (\ref{acca1})-(\ref{acca2}), that this
perturbation can be either due to the addition of $N$ particles, or
due to a connectivity shift, and leads to a linear response of the
free energy.  Hence the theorem above combines two invariance
properties of the optimal RaMOSt measure $\Omega$: the one of the
cavity part with respect to the cavity dynamics, and the one of the
fugacity part with respect to connectivity shifts.  The invariance
with respect to the cavity dynamics is a special case of
\emph{Quasi-Stationarity}, i.e. the invariance up to a correcting
factor under random shifts (see \cite{ar,ass2} for a detailed
introduction). The Parisi ultrametric Ansatz, both for dilute and for
fully connected Gaussian models \cite{talabook}, is based on
Hierarchical Random Probability Cascades, which exhibit the
Quasi-Stationarity of the Generalized Random Energy Model \cite{ass2,
  talabook}. A very stimulating conjecture is that Random Probability
Cascades include all the quasi-stationary structures.

As an aside remark, we point out that for stochastically stable
systems, a dynamical order parameter can be defined and related to the
static order parameter.  Ultrametricity in the dynamics implies static
ultrametricity, which is in turn implied by so-called
\textit{separability}, and connected to the idea of \textit{overalp
  equivalence}.  We refer to \cite{fmpp,prt,ldsf} for details, here we
just wish to stress that all these concepts and the one of
self-averaging are intimately related, and have deep physical
meanings.


\section{Non-negativity of the average of
  multi-overlaps}\label{positivi}

We know from \cite{lds2,gt1} that
above the critical temperature $\beta_{c}$ all the multi-overlaps
(including the 2-overlap) are identically zero, as the replica
symmetric solution holds.
We also know, from \cite{gt1}, that the (rescaled) 2-overlap
shows diverging fluctuations at the critical temperature
$\beta_{c}$
where the replica symmetry is broken,
while the (rescaled) multi-overlaps of more than two replicas do not
exhibit diverging fluctuations at this inverse temperature.
If the expression of the fluctuations of the rescaled multi-overlaps
given in \cite{gt1} could be proven to be valid down
to suitable lower temperatures, we would have what is 
physically a common belief \cite{vb}, i.e. that
the critical temperature
$\beta^{(2n)}_{c}$ at which
the fluctuations of $\sqrt{N}q_{2n}$ diverge is given by
$$
\tanh^{2n}(\beta^{(2n)}_{c})=\frac{1}{2\alpha}\ , \ \alpha>\frac12\ .
$$
So that $q_{2n}$ would be zero up to $\beta_{c}^{(2n)}$,
where it would start its concave increase toward 1 as $\beta\to\infty$.

We want to show that
the 2-overlap exhibit the same phase transition as
in the Gaussian SK model \cite{abds}:
if we apply an external field and then remove it in the
thermodynamic limit, the 2-overlap remains strictly
positive below its critical temperature, where its
variance becomes non-zero.
The same is expected to hold for all multi-overlaps.

Let us introduce the following notation:
$\omega_{\bar{\alpha}}(\cdot)\ ,\ \langle \cdot \rangle_{\bar{\alpha}}$
denote the usual
expectations except for a perturbation in
the {\sl Boltzmannfaktor}, which is assumed here to be
$$
\exp\left[-\beta \bigg(H_{N}(\sigma;\alpha)-\sum_{i=1}^{N}
\bigg(\sum_{\nu=1}^{P^{i}_{\bar{\alpha}}}
\tilde{J}^{i}_{\nu}\bigg)\sigma_{i}\bigg)\right]\ ,
$$
i.e. the initial {\sl Boltzmannfaktor} is
perturbed with independent copies of an external field
$\tilde{h}(\bar{\alpha})=\sum_{\nu=1}^{P_{\bar{\alpha}}}
\tilde{J}_{\nu}$
modulated by $\bar{\alpha}$.

This section is devoted to the next
\begin{theorem}\label{positive} The following holds:
\begin{enumerate}
\item for any inverse temperature $\beta$
$$
\lim_{N\to\infty}\langle q_{12}
\rangle_{\bar{\alpha}=2\alpha/N} \geq 0\ ;
$$
\item for any $\beta>\beta^{(2)}_{c}$,
defined by $2\alpha\tanh^{2}(\beta^{(2)}_{c})=1$,
$$
\lim_{N\to\infty}\langle q_{12}
\rangle_{\bar{\alpha}=2\alpha/N}>0\ .
$$
\end{enumerate}
\end{theorem}
We will see that our method would imply immediately
the analogous statement for all multi-overlaps, should 
the formula for their fluctuation be proven to hold at lower 
temperatures as well.

The theorem will be a simple consequence of two lemmas,
which require a definition as well.
\begin{lemma}\label{n+1}
Consider the set of indices
$\{i_1,..,i_r\}$, with $r\in [1,N]$. Then
$$
\lim_{\bar{\alpha} \to
2\alpha/N}\omega_{N, \bar{\alpha}}(\sigma_{i_1}\cdots\sigma_{i_r})=
\omega_{N+1}(\sigma_{i_1}\cdots\sigma_{i_r}\sigma_{N+1}^r)
+O(\frac{1}{N} )\ ,
$$
where $r$ is an exponent (not a replica index) and
we have made explicit the dependence of $\omega$
on the size of the system.
\end{lemma}
This lemma is a consequence of the fact that with the chosen
$\bar{\alpha}$ the presence of the external field is equivalent to
the introduction of an additional particle, labelled $N+1$. This
should be clear from (\ref{step})-(\ref{acca1}) and from the gauge
symmetry with respect to the transformation
$\sigma_{i}\to\sigma_{i}\sigma_{N+1}$, but the reader may want to
refer to Lemma \ref{n+1} in \cite{abds} for details. We do not need here
the most general version of the formalism presented there, and
limit ourselves to the following
\begin{definition}
A polynomial function of some overlaps is called
\begin{itemize}
\item {\bf filled} if every replica appears an even number of 
times in it;
\item {\bf fillable} if it can be made filled by multiplying 
it by exactly one
multi-overlap of appropriately chosen replicas;
\end{itemize}
\end{definition}
The next lemma is a consequence
of the previous one, where the exponent $r$ 
is always even for filled polynomials.
\begin{lemma}\label{media}
In the $N \to \infty$ limit the
average $\langle \cdot \rangle_{\bar{\alpha}}$
of filled polynomials is not affected by the
presence of the perturbation
modulated by $\bar{\alpha}$, that is, for instance,
$$
\int_{\bar{\alpha}_1}^{\bar{\alpha}_2}\langle
q_{12}q_{23}q_{13}\rangle_{\bar{\alpha}}
d\bar{\alpha}=
\int_{\bar{\alpha}_1}^{\bar{\alpha}_2}\langle 
q_{12}q_{23}q_{13}\rangle
d\bar{\alpha}\ \ ,
$$
for any interval $[\bar{\alpha}_1,\bar{\alpha}_2]$.
\end{lemma}
Recall that the Gibbs measure in case of several
replicas is the product measure, and that
the perturbation given by $\bar{\alpha}$
is equivalent to the addition of a new
spin to the system, up to neglibile 
terms vanishing in the thermodynamic limit. 
So if we consider in Lemma \ref{n+1}
for a filled polynomial of replicas, then 
we have the product of several Gibbs 
expectations, in each of which the added spin
appears with the exponent $r$ even, being thus
uneffective.

We are not going to use this lemma in its full generality, we 
will use it only in the case where the filled
polynomial is a squared multi-overlap. 
At least in this case, as the free energy is a series with
averaged squared multi-overlaps, the regularity of the
free energy is equivalent to the fact that the pertubed
averages of squared multi-overlaps tend to the unperturbed ones.
This is discussed in \cite{ac,lds1,g1} but also in 
Lemma \ref{lemma} from next section, and the key concept
is always the convexity of the free energy in the connectivity.

We will refer to statements, like the one in the Lemma above,
true only taking the integral,
as valid in $\bar{\alpha}$-average, and we will omit to write
the integral.
From the previous lemma we will deduce the next
\begin{proposition}\label{pieni}
Let $Q_{1\cdots 2n}$ be a fillable polynomial of the
overlaps, such that $q_{1\cdots 2n}Q_{1\cdots 2n}$ is filled.
Then
$$
\lim_{N \to \infty}
\langle Q_{1\cdots 2n}\rangle_{\bar{\alpha}=2\alpha/N}
= \langle q_{1\cdots 2n}Q_{1\cdots 2n} \rangle\ ,
$$
where the right hand side is understood to be evaluated in the
thermodynamic limit.
\end{proposition}
\textbf{Proof}. Let us assume for a generic overlap correlation
function $Q$, of $s$ replicas, the following representation $$
Q =
\prod_{a=1}^s\sum_{i_l^a}\prod_{l=1}^{n^a}\sigma_{i_l^a}^a I(\{i_l^a
\}) $$
where $a$ labels the replicas, the internal productory takes
into account the spins (labelled by $l$) which contribute to the
a-part of the overlap $q_{a,a'}$ and runs to the number of time that
the replica $a$ appears in $Q$, the external productory takes into
account all the contributions of the internal one and the $I$ factor
fixes the constraints among different replicas in $Q$; so, for
example, $Q=q_{12}q_{23}$ can be decomposed in this form noting that
$s=3$, $n^1=n^3=1,n^2=2$,
$I=N^{-2}\delta_{i_1^1,i_1^3}\delta_{i_1^2,i_2^3}$, where the $\delta$
functions fixes the links between replicas $1,2 \rightarrow q_{1,2}$
and $2,3 \rightarrow q_{2,3}$.  The averaged overlap correlation
function is 
$$
\langle Q \rangle =\sum_{i_l^a}I(\{i_l^a
\})\prod_{a=1}^s \omega_{\alpha}(\prod_{l=1}^{n^a}\sigma_{i_l^a}^a)\ .
$$
Now if $Q$ is a fillable polynomial, and we evaluate it at
$\bar{\alpha}=2 \alpha / N$, let us decompose it, using the
factorization of the $\omega$ state on different replica, as 
$$
\langle Q \rangle =\sum_{i_l^a,i_l^b}I(\{i_l^a \}, \{i_l^b
\})\prod_{a=1}^u \omega_a ( \prod_{l=1}^{n^a}\sigma_{i_l^a}^a)
\prod_{b=u}^s \omega_b ( \prod_{l=1}^{n^b}\sigma_{i_l^b}^b),
$$
where
$u$ stands for the number of the unfilled replicas inside the
expression of $Q$. So we split the measure $\Omega$ into two different
subset $\omega_{a}$ and $\omega_{b}$: in this way the replica
belonging to the $b$ subset are always in even number, while the ones
in the $a$ subset are always odds. Applying the gauge 
$\sigma_i^a\rightarrow \sigma_i^a\sigma_{N+1}^a, 
\forall i \in (1,N)$ 
the even
measure is unaffected by this transformation $(\sigma_{N+1}^{2n}
\equiv 1)$ while the odd measure takes a $\sigma_{N+1}$ inside the
Boltzmann measure (Lemma 1,2).  $$
\langle Q \rangle =
\sum_{i_l^a,i_l^b}I(\{i_l^a \}, \{i_l^b \}) \prod_{a=1}^u \omega (
\sigma_{N+1}^a \prod_{l=1}^{n^a}\sigma_{i_l^a}^a) \prod_{b=u}^s \omega
( \sigma_{N+1}^b\prod_{l=1}^{n^b}\sigma_{i_l^b}^b)$$
At the end we can
replace in the last expression the subindex $N+1$ of $\sigma_{N+1}$ by
$k$ for any $k \neq \{ i_l^a \}$ and multiply by one as
$1=N^{-1}\sum_{k=0}^N$.  Up to orders $O(1/N)$, which go to zero in
the thermodynamic limit, we have the proof $\Box$.

At this point the first part of Theorem \ref{positive} is simply a
corollary of this proposition in the case $Q=q$, and of the fact
that all multi-overlaps (including the 2-overlap) 
are zero above the critical temperature $1/\beta_{c}^{(2)}$,
and the 2-overlap starts fluctuating below, so that $\langle
q^2_{2}\rangle$ is strictly positive and coincides (in the limit)
with $\langle q_{2}\rangle$ by Proposition \ref{pieni}.
The fact that $\langle q^{2}_{2}\rangle$ is strictly positive
for $\beta>\beta_{c}^{(2)}$ is obvious from the fact that
the replica symmetric solution does not hold in this region \cite{gt1}.

We will discuss the generalization of Theorem \ref{positive} to 
multi-overlaps in \cite{bdsfg}.


\section{Stability relations from Quasi-Stationarity}
\label{qs}

In this section we want to prove the following
\begin{theorem}\label{acmain} The consequences of stochastic
stability in fully connected models extend
to dilute spin glasses, and constraints analogous to
those found for overlaps of two replicas only
hold for multi-overlaps. More precisely,
\begin{enumerate}
\item the Aizenman-Contucci relations hold
in dilute spin glasses. A first example is
$$
\langle q^{2}_{12}q^{2}_{13}\rangle=\frac14\langle q^{4}_{12}\rangle
+\frac34\langle q^{2}_{12}q^{2}_{34}\rangle\ ;
$$
\item further relations for multi-overlaps hold
in dilute spin glasses. A first example is
$$
\langle q^{2}_{1234}q^{2}_{15}\rangle
=\frac{3}{8}\langle q^{2}_{1234}q^{2}_{12}\rangle
+\frac{5}{8}\langle q^{2}_{1234}q^{2}_{56}\rangle\ .
$$
\end{enumerate}
\end{theorem}

We start addressing the proof of the theorem by proving a lemma
that gives the explicit form of the contribution
to the free energy of a connectivity shift.
\begin{lemma}\label{lemma}
Let $\Omega\ ,\ \langle\cdot\rangle$ be the usual Gibbs
and quenched Gibbs expectations at inverse temperature
$\beta$, associated with the
Hamiltonian $H_{N}(\sigma, \alpha; \mathcal{J})$. Then,
in the whole region where the parameters are uniquely defined
\begin{equation}\label{stability}
\lim_{N\to\infty}\mathbb{E}\ln\Omega\exp\bigg(\beta^{\prime}
\sum_{\nu=1}^{P_{\alpha^{\prime}}}
J^{\prime}_{\nu}\sigma_{i^{\prime}_{\nu}}
\sigma_{j^{\prime}_{\nu}}\bigg)
=\alpha^{\prime}\sum_{n=1}^{\infty}\frac{1}{2n}
\tanh^{2n}(\beta^{\prime})(1-\langle q^{2}_{2n}\rangle)\ ,
\end{equation}
where the random
variables $P_{\alpha^{\prime}}, \{J^{\prime}_{\nu}\}$,
$\{i^{\prime}_{\nu}\},\{j^{\prime}_{\nu}\}$ are independent
copies of the analogous random variables
appearing in the Hamiltonian in $\Omega$.
\end{lemma}
Notice that, in distribution
\begin{equation}
\label{continuo1}
\beta\sum_{\nu=1}^{P_{\alpha N}}
J_{\nu}\sigma_{i_{\nu}}\sigma_{j_{\nu}}+\beta^{\prime}
\sum_{\nu=1}^{P_{\alpha^{\prime}}}
J^{\prime}_{\nu}\sigma_{i^{\prime}_{\nu}}\sigma_{j^{\prime}_{\nu}}
\sim\beta\sum_{\nu=1}^{P_{(\alpha+\alpha^{\prime}/N)N}}
J^{\prime\prime}_{\nu}\sigma_{i_{\nu}}\sigma_{j_{\nu}}
\end{equation}
where $\{J^{\prime\prime}_{\nu}\}$ are independent copies
of $J$ with probability $\alpha N/(\alpha N+\alpha^{\prime})$ and
independent copies of $J\beta^{\prime}/\beta$ with probability
$\alpha^{\prime}/(\alpha N+\alpha^{\prime})$. In the right hand
side above, the quenched random variables will be collectively denoted
by $\mathcal{J}^{\prime\prime}$.
Notice also that
the sum of Poisson random variables
is a Poisson random variable with
mean equal to the sum of the means, and hence we can write
\begin{equation}
  \label{continuo2}
  A_{t}\equiv\mathbb{E}\ln\Omega\exp\bigg(\beta^{\prime}
  \sum_{\nu=1}^{P_{\alpha^{\prime}t}}
  J^{\prime}_{\nu}\sigma_{i^{\prime}_{\nu}}
  \sigma_{j^{\prime}_{\nu}}\bigg)
  =\mathbb{E}\ln\frac{Z_{N}(\alpha_{t};
    \mathcal{J}^{\prime\prime})}{Z_{N}(\alpha;\mathcal{J})}\ ,
\end{equation}
where we defined, for $t\in[0, 1]$,
\begin{equation}
  \label{continuo3}
  \alpha_{t}=\alpha+\alpha^{\prime}\frac{t}{N}
\end{equation}
so that $\alpha_{t}\rightarrow\alpha\ \forall \ t$ as $N\to\infty$.\\
\textbf{Proof}.
Let us compute the $t$-derivative of $A_{t}$, as defined in
(\ref{continuo2})
\begin{equation*}
  \frac{d}{dt}A_{t}=
  \mathbb{E}\sum_{m=1}^{\infty}\frac{d}{dt}\pi_{\alpha^{\prime} t}(m)
  \ln\sum_{\sigma}\exp\bigg(\beta^{\prime}\sum_{\nu=1}^{m}
  J^{\prime}_{\nu}\sigma_{i^{\prime}_{\nu}}
  \sigma_{j^{\prime}_{\nu}}\bigg)\ .
\end{equation*}
Using the following
elementary property of the Poisson measure
\begin{equation}\label{poisson}
  \frac{d}{dt}\pi_{t\zeta}(m)=\zeta(\pi_{t\zeta}(m-1)-\pi_{t\zeta}(m))
\end{equation}
we get
\begin{eqnarray*}
  \frac{d}{dt}A_{t}&=&
  \alpha^{\prime}\mathbb{E}\sum_{m=0}^{\infty}
  [\pi_{\alpha^{\prime} t}(m-1)
  -\pi_{\alpha^{\prime} t}(m)]
  \ln\sum_{\sigma}\exp(\beta^{\prime}\sum_{\nu=1}^{m}J^{\prime}_{\nu}
  \sigma_{i^{\prime}_{\nu}}\sigma_{j^{\prime}_{\nu}})\\
  {}&=&\alpha^{\prime}\mathbb{E}\ln\sum_{\sigma}
  \exp(\beta^{\prime} J^{\prime}\sigma_{i^{\prime}_{m}}
  \sigma_{j^{\prime}_{m}})
  \exp(\beta^{\prime}\sum_{\nu=1}^{P_{\alpha^{\prime} t}}
  J^{\prime}_{\nu}
  \sigma_{i^{\prime}_{\nu}}\sigma_{j^{\prime}_{\nu}})\\
  {}&{}&\hspace{3cm}-\alpha^{\prime}\mathbb{E}\ln\sum_{\sigma}
  \exp(\beta^{\prime}\sum_{\nu=1}^{P_{\alpha^{\prime} t}}
  J^{\prime}_{\nu}
  \sigma_{i^{\prime}_{\nu}}\sigma_{j^{\prime}_{\nu}})\\
  {}&=&\alpha^{\prime}\mathbb{E}\ln\Omega_{t}
  \exp(\beta^{\prime} J^{\prime}\sigma_{i^{\prime}_{m}}
  \sigma_{j^{\prime}_{m}})\ ,
\end{eqnarray*}
where we included the $t$-dependent weights
in the average $\Omega_{t}$.
Now use the following identity
$$
\exp(\beta^{\prime} J^{\prime}\sigma_{i}\sigma_{j})
=\cosh(\beta^{\prime} J^{\prime})+\sigma_{i}\sigma_{j}
\sinh(\beta^{\prime} J^{\prime})
$$
to get
\begin{equation*}
  \frac{d}{dt}A_{t}
  =\alpha^{\prime}\mathbb{E}\ln\Omega_{t}
  [\cosh(\beta^{\prime} J^{\prime})(1+\tanh(\beta^{\prime}
  J^{\prime})\sigma_{i^{\prime}_{m}}
  \sigma_{j^{\prime}_{m}})]\ .
\end{equation*}
It is clear that
\begin{equation*}
  \mathbb{E}\ \omega_{t}^{2n}(\sigma_{i_{m}}\sigma_{j_{m}})
  =\langle q^{2}_{2n}\rangle_{t}\ ,
\end{equation*}
so we now expand the logarithm in power series and see that,
in the limit of large $N$, as $\alpha_{t}\to\alpha$
the result does not depend on $t$,
everywhere the measure $\langle\cdot\rangle_{t}$
is continuous
as a function of the parameter $t$.
From the comments that preceded
the current proof, formalized
in (\ref{continuo1})-(\ref{continuo2})-(\ref{continuo3}),
this is the same as assuming that
$\Omega$ is regular as a function of $\alpha$,
because
$J^{\prime\prime}\to J$ in the sense that in the large
$N$ limit $J^{\prime\prime}$ can only take
the usual values $\pm 1$ since the probability of being
$\pm \beta^{\prime}/\beta$ becomes zero.
Therefore integrating
over $t$ from 0 to 1 is the same as multiplying by 1.
Due to the symmetric distribution of $J$,
the expansion of the logarithm yields the right hand side of
(\ref{stability}),
where the odd powers are missing. $\Box$
\begin{remark}
The same result
holds for the hierarchical Parisi trial structure
\cite{lds2}. In general, what we study in this section
relies only on (\ref{stability}), that holds for quasi-stationary
RaMOSt's, to which our results therefore extend.
\end{remark}
\textbf{Proof of Theorem \ref{acmain}}. Consider
once again
\begin{equation}\label{defhath}
\hat{H}=-\sum_{\nu=1}^{P_{\alpha^{\prime}}}
J^{\prime}_{\nu}\sigma_{i^{\prime}_{\nu}}\sigma_{j^{\prime}_{\nu}}\ .
\end{equation}
We will let again $\Omega$ be the infinite volume Gibbs measure
associated with the VB Hamiltonian at connectivity $\alpha$
and inverse temperature $\beta$.

Due to the symmetry of $\Omega$,
we have \cite{bds}
$$
\mathbb{E}\ln\Omega\exp(\beta^{\prime} \hat{H})
=\frac{1}{2}\mathbb{E}\ln\Omega\exp(-\beta^{\prime}
 (\hat{H}-\hat{H}^{\prime}))\ ,
$$
where we ``replicated'' $\hat{H}^{\prime}=\hat{H}(\sigma^{\prime})$.

For the left hand side of (\ref{stability}), a tedious expansion yields
\begin{eqnarray*}
  && \frac{1}{2}\ln\Omega\exp(-\beta^{\prime}
  (\hat{H}-\hat{H}^{\prime}))=\\
  && \quad \beta^{\prime 2}\frac{1}{4}[2\Omega(\hat{H}^{2})
  -2\Omega^{2}(\hat{H})] +\\
  && \quad\beta^{\prime 4} \frac{1}{24}[\Omega(\hat{H}^{4})
  -4\Omega(\hat{H}^{3})\Omega(\hat{H})-3\Omega^{2}(\hat{H}^{2})
  +12\Omega(\hat{H}^{2})\Omega^{2}(\hat{H})
  -6\Omega^{4}(\hat{H})] +\\
  && \quad \beta^{\prime 6}[\frac{1}{6!}\Omega(\hat{H}^{6})-
  \frac{1}{5!}\Omega(\hat{H}^{5})\Omega(\hat{H})-
  \frac{1}{48}\Omega(\hat{H}^{4})\Omega(\hat{H}^{2})
  -\frac{1}{72}\Omega^{2}(\hat{H}^{3})\\
  &&\qquad+\frac{1}{6}\Omega(\hat{H})
  \Omega(\hat{H}^{2})\Omega(\hat{H}^{3})+
  \frac{1}{24}\Omega^{3}(\hat{H}^{2})
  +\frac{1}{24}\Omega^{2}(\hat{H})\Omega(\hat{H}^{4})
  -\frac{1}{6}\Omega(\hat{H}^{3})\Omega^{3}(\hat{H})\\
  &&\qquad -\frac{3}{8}\Omega^{2}(\hat{H})\Omega^{2}(\hat{H}^{2})+
  \frac{1}{2}\Omega(\hat{H}^{2})\Omega^{4}(\hat{H})
  -\frac{1}{6}\Omega^{6}(\hat{H})]+O(\beta^{\prime 8})\ ,
\end{eqnarray*}
of which we have to take the quenched expectation $\mathbb{E}$,
using the formulas in Appendix \ref{formulas}.
For the right hand side of (\ref{stability}),
the expansion of the hyperbolic tangent,
performed explicitly for convenience in 
Appendix \ref{formulas}, leads to
\begin{eqnarray*}
  &&\alpha^{\prime}\sum_{n>0}\frac{1}{2n}\tanh^{2n}(\beta^{\prime})
  (1-\langle q^{2}_{1\cdots 2n}\rangle)  = \\
  &&\hspace{3cm}\beta^{\prime 2}\alpha^{\prime}(\frac{1}{2}
  -\frac{1}{2}\langle q^{2}_{12}\rangle)+\\
  && \hspace{3cm} \beta^{\prime 4}\alpha^{\prime}(-\frac{1}{12}
  +\frac{1}{3}\langle q^{2}_{12}\rangle
  -\frac{1}{4}\langle q^{2}_{1234}\rangle)+\\
  && \hspace{3cm}\beta^{\prime 6}\alpha^{\prime}(\frac{2}{90}-
  \frac{17}{90}\langle q^{2}_{12}\rangle
  +\frac{1}{3}\langle q^{2}_{1234}\rangle-
  \frac{1}{6}\langle q^{2}_{123456}\rangle)+O(\beta^{\prime 8})
\end{eqnarray*}

Recall that the averages $\langle\cdot\rangle$ do
not depend on $\beta^{\prime}$, so we now have two
power series in $\beta^{\prime}$ that we can
equate term by term.
The order zero and the odd orders are absent on both sides.
Let us consider the second order.
Taking into account the formulas
given in Appendix \ref{formulas}, we have two
identical (constant) monomials in $\alpha^{\prime}$
$$
\frac12(1-\langle q^{2}_{12}\rangle)=
\frac12(1-\langle q^{2}_{12}\rangle)
$$
and we gain no information.
Let us move on to
order four: using again the formulas
given in Appendix \ref{formulas}, we have the equality of two
polynomials of degree two in $\alpha^{\prime}$. 
There is no constant term,
and no second power in the right hand side. 
Equating term by term we get
a trivial identity for the linear part in
$\alpha^{\prime}$ (and hence no information).
From the quadratic term in $\alpha^{\prime}$, on the other hand,
we obtain the following relation
\begin{equation}
  \label{primaac}
  \langle q^{2}_{12}q^{2}_{13}\rangle=\frac14\langle q^{4}_{12}\rangle
  +\frac34\langle q^{2}_{12}q^{2}_{34}\rangle
\end{equation}
which is the first AC relation we wanted to prove.

Notice
that in the linear part in $\alpha^{\prime}$ a
multi-overlap is present: $q_{1234}$,
but it cancels. By contrast, in the quadratic part
only 2-overlaps are present.

Let us now focus on the sixth order. Again we get a
useless identity from the linear part
in $\alpha^{\prime}$, from which 4-overlaps and
6-overlaps get canceled out, and we
gain no information.
From the quadratic part, where 4-overlaps are
present (but no 6-overlaps are there),
$\langle q^{2}_{12}\rangle$ disappears, while the other
terms with 2-overlaps only cancel because of the previous
AC relation. In fact, from the expansion we have
\begin{multline*}
  \frac{15}{6!}-\frac{15}{5!}\langle q^{2}_{12}\rangle-
\frac{7}{48}-\frac{1}{6}
  \langle q^{4}_{12}\rangle-\frac{15}{72}\langle q^{2}_{12}\rangle
  +\frac{7}{6}\langle q^{2}_{12}\rangle+\\
  \frac{4}{3}\langle q^{2}_{12}q^{2}_{23}\rangle+\frac{1}{8}
  +\frac{1}{2}\langle q^{4}_{12}\rangle+\frac{1}{24}\langle 
q^{2}_{12}\rangle
  +\frac{1}{3}\langle q^{2}_{12}q^{2}_{23}\rangle
  +\frac{1}{4}\langle q^{2}_{12}\rangle-\\
  \frac{1}{2}\langle q^{2}_{1234}\rangle-2\langle 
q^{2}_{12}q^{2}_{34}\rangle
  -\frac{9}{8}\langle q^{2}_{12}\rangle-3\langle 
q^{2}_{12}q^{2}_{23}\rangle-
  \frac{3}{2}\langle q^{2}_{12}q^{2}_{1234}\rangle+\\
  \frac{1}{2}\langle q^{2}_{1234}\rangle+3\langle 
q^{2}_{12}q^{2}_{34}\rangle
  +4\langle q^{2}_{12}q^{2}_{2345}\rangle-
  \frac{15}{6}\langle q^{2}_{12}q^{2}_{3456}\rangle=0\ .
\end{multline*}
So we are left, after a few trivial calculations,
with the following new relation for multi-overlaps
\begin{equation}
  \label{secondaac}
  \langle q^{2}_{1234}q^{2}_{15}\rangle
  =\frac{3}{8}\langle q^{2}_{1234}q^{2}_{12}\rangle
  +\frac{5}{8}\langle q^{2}_{1234}q^{2}_{56}\rangle
\end{equation}
announced in the statement of the theorem.

In the cubic part in $\alpha^{\prime}$
only overlaps of two replicas are present,
and only the monomials of order six remain, as the ones of
lower degree cancel out directly. The remaining relation is
\begin{multline*}
\frac{1}{12}\langle q^{6}_{12}\rangle-\langle 
q^{2}_{12}q^{4}_{23}\rangle
+\frac{3}{4}\langle q^{2}_{12}q^{4}_{34}\rangle-\frac{1}{3}
\langle q^{2}_{12}q^{2}_{23}q^{2}_{31}\rangle+\\
\langle q^{2}_{12}q^{2}_{13}q^{2}_{14}\rangle+
3\langle q^{2}_{12}q^{2}_{23}q^{2}_{34}\rangle
-6\langle q^{2}_{12}q^{2}_{23}q^{2}_{45}\rangle
+\frac{5}{2}\langle q^{2}_{12}q^{2}_{34}q^{2}_{56}\rangle=0\ .
\end{multline*}
Proceeding further, the expansion generates
all the identities due to stochastic stability,
in full agreement with the self-averaging
identities found in \cite{flt,ldsf}. $\Box$

We will provide a more general and systematic
form of the relations in the next section.


\section{Stability relations from the cavity streaming
  equation}\label{barra}

In this section we study a family of constraints on the
distribution of the overlaps. To address this task we
will consider the quenched expectation of a generic function of
$s$ replicas, with respect to the perturbed measure
with weights
\begin{equation}\label{pesi}
  \exp\left(-\beta H_{N}(\sigma;\alpha)+\beta^{\prime}
    \sum_{\nu=1}^{P_{2\alpha^{\prime} t}}
    \tilde{J}_{\nu}\sigma_{i_{\nu}}\right)\ ,
\end{equation}
whose use will be indicated with a subscript $t$ in the expectations.
Once again, $\{\tilde{J}_{\nu}\}$ are independent copies of $J$.

From now on let us put $\theta = \tanh(\beta^{\prime})$,
and, assuming $J=\pm 1$, we have
$\theta^{2n}= \mathbb{E}\tanh^{2n}(\beta^{\prime} J)$,
$\tanh^{2n+1}(\beta^{\prime}J)=J\theta^{2n+1}$
$\forall\  n \in \mathbb{N}$.
Let us also just put $\omega_{t}=\omega_{t}(\sigma)$, with a slight
abuse of notation.
\begin{proposition}\label{derivata}
  Let $\Phi$ be a function of $s$ replicas. Then
  the following cavity streaming equation holds
  \begin{multline}\label{stream}
    \frac{d\langle \Phi \rangle_{t}}{dt} =
    -2\alpha^{\prime}\langle \Phi
    \rangle_t +2\alpha^{\prime} \mathbb{E}[\Omega_{t}
    \Phi \{ 1 + J\sum_{a}^{1,s}
    \sigma^{a}_{i_{1}}\theta
    + \sum_{a < b}^{1,s}
    \sigma^{a}_{i_{1}}\sigma^{b}_{i_{1}} \theta^2 + \\
    J\sum_{a < b < c}^{1,s}
    \sigma^{a}_{i_{1}}\sigma^{b}_{i_{1}}\sigma^{c}_{i_{1}}
    \theta^3 + \cdots \} \{ 1 - s
    J\theta\omega_{t} + \frac{s(s+1)}{2!}\theta^2
    \omega_{t}^{2} - \\
    -\frac{s(s+1)(s+2)}{3!}J\theta^3\omega_{t}^{3}
    + \cdots \}]\ \ \forall\ \theta
  \end{multline}
\end{proposition}
\textbf{Proof}.
Let us explicitly perform the calculation of the derivative,
using (\ref{poisson}).
\begin{eqnarray*}
  \frac{d}{dt}
  \mathbb{E}Z_t^{-1} \sum_{\sigma} \Phi \exp({
    \sum_{a=1}^s(\beta\sum_{\nu}^{P_{\alpha^{\prime} N}}
    J_{\nu}\sigma_{i_{\nu}}^{a}\sigma_{j_{\nu}}^{a}+\beta^{\prime}
    \sum_{\nu}^{P_{2\alpha^{\prime} t}}
    \tilde{J}_{\nu}\sigma_{i_{\nu}}^{a}})) =\\
  2\alpha^{\prime}\mathbb{E}\frac{\Omega_{t}[\Phi
    \exp(\beta^{\prime}\sum_{a=1}^s J
    \sigma^{a}_{i_{1}})]}{\Omega_{t}
    \exp(\beta^{\prime}\sum_{a=1}^s J \sigma^{a}_{i_{1}})}
  - 2\alpha^{\prime}\langle \Phi \rangle_t =\\
  2\alpha^{\prime}\mathbb{E} \frac{\Omega_{t}[\Phi
    \prod_{a=1}^s (\cosh(\beta^{\prime} J) +
    \sigma^{a}_{i_{1}} \sinh(\beta^{\prime} J))]}{\Omega_{t}
    [\prod_{a=1}^s
    (\cosh(\beta^{\prime} J) + \sigma^{a}_{i_{1}}
    \sinh(\beta^{\prime} J))]}
  - 2\alpha^{\prime}\langle
  \Phi \rangle_t =\\
  2\alpha^{\prime}\mathbb{E}
  \frac{\Omega_{t}[\Phi\prod_{a}^s(1+J\theta\sigma^{a}_{i_{1}})]}{(1
    +\theta\omega)^s}-2\alpha^{\prime}\langle \Phi\rangle_t\ .
\end{eqnarray*}
Now note that
\begin{multline*}
  \frac{1}{(1+J\theta\omega_{t})^s}=1- J s
  \theta\omega_{t}+
  \frac{s(s+1)}{2!}\theta^{2}\omega^2_{t}-\\
  J\frac{s(s+1)(s+2)}{3!}\theta^3\omega^3_{t}
  +\frac{s(s+1)(s+2)(s+3)}{4!}\theta^{4}\omega^4_{t}-\cdots
\end{multline*}
and that
$$
\prod_{a=1}^s(1+J\theta\sigma^a_{i_1})=1+
J\sum_a^{1,s}\sigma^a_{i_1} \theta+
\sum_{a<b}^{1,s}\sigma_{i_1}^{a}\sigma_{i_1}^{b}\theta^2+
J\sum_{a<b<c}^{1,s}\sigma_{i_1}^{a}
\sigma_{i_1}^{b}\sigma_{i_1}^{c}\theta^3 +\cdots\ .
$$
The theorem follows immediately. $\Box$

In the limit $\alpha^{\prime}\to\infty$ as
$\beta^{\prime}_{SK}=2\alpha\theta^{2}$
is kept constant, the
powers of $\theta$ higher than two are killed, and we recover
the equation for the Gaussian SK model \cite{bds}
\begin{equation}
  \label{generale2}
  \frac{d\langle \Phi \rangle_{t}}{dt}=  \langle \Phi(\sum_{a <
    b}^{1,s}q_{ab} -s \sum_{a}^{1,s} q_{a, s+1} +
  \frac{s(s+1)}{2}q_{s+1,s+2}) \rangle_t\ .
\end{equation}

If in the previous theorem we take a function $\Phi$ whose average does
not depend on $t$, the left hand side of (\ref{stream}) is zero,
and we have a polynomial in $\theta$ (and hence in $\beta^{\prime}$)
on the right
hand side that can be equated to zero term by term (we
do not need to re-expand in $\beta^{\prime}$
and equate the new coefficients to zero).
We know from Lemma \ref{media} that the left
hand side of (\ref{stream}) does not depend on $t$ when,
for instance,
$\Phi$ is a filled polynomial. If, in each term
of the expansion that we equate to zero in this case, we additionally
take $t=1, \alpha^{\prime}=\alpha , \beta^{\prime}=\beta$,
and let $N\to\infty$, then we also
guarantee that the fillable polynomials get filled, thanks
to Proposition \ref{pieni}. In other words
\begin{proposition}
  The generator of the constraints on the distribution
  of the overlap is:
  $$
  \lim_{N\to\infty}
  \partial_t \langle \Phi \rangle_t =
  0
  $$
  where $\Phi$ is filled, and $t=1,\alpha^{\prime}=\alpha,
  \beta^{\prime}=\beta$.
\end{proposition}
Let us consider the first simple example,
$\Phi=q_{12}^{2}$. Proposition \ref{derivata} then yields
\begin{multline*}
  \lim_{N\to\infty}
  \partial_{t}\langle q_{12}^2 \rangle_t =
  \lim_{N\to\infty}
  \langle q_{12}^3 - 4 q_{12}^2q_{23} + 3 q^2_{12}q_{34}
  \rangle_t\theta^{2} +O(\theta^{4})=\\
  \lim_{N\to\infty}
  \langle q_{12}^3 - 4 q_{12}^2q_{23} + 3 q^2_{12}q_{34}
  \rangle_t\beta^{\prime 2} +O(\beta^{\prime 4})=0\\
  \!\!\! \Rightarrow \ \ \
  \langle q_{12}^4 - 4 q_{12}^2q_{23}^2 + 3 q_{12}^2
  q_{34}^2 \rangle=0\ ,
\end{multline*} which is understood to be taken
in the thermodynamic limit and is just the Aizenman-Contucci
relation we have found already in the previous section.

If we choose instead $\Phi=q^{2}_{1234}$, we obtain
\begin{multline*}
  \lim_{N\to\infty}\partial_{t}\langle q_{1234}^2 \rangle_t =
  \langle \theta^2(3q_{1234}^2q_{12}^2 - 8q_{1234}^2q_{15}^2 + 10
  q_{1234}^2q_{15^2})+\\
  \theta^4(q_{1234}^4-16q_{1234}^2q_{1235}^2+60q_{1234}^2q_{1256}^2
  -80q_{1234}^2q_{1567}^2+35q^2_{1234}q^2_{5678})\rangle
  +O(\theta^{6})=0
\end{multline*}
From the order two in $\beta^{\prime}$ we have the coefficient of
$\theta^{2}$, and equating to zero gives 
$$
\langle q_{1234}^2q_{15}^2
\rangle = \frac{3}{8}\langle q_{1234}^2 q_{12}^2\rangle + \frac{5}{8}
\langle q_{1234}^2q_{56}^2 \rangle\ .  
$$
At the order four in
$\beta^{\prime}$ we have leftover term from the order two, which thus
vanish, and the coefficient of $\theta^{4}$: 
$$
\langle q_{1234}^4
\rangle = \langle 16q_{1234}^2q_{1235}^2 -60q_{1234}^2q_{1256}^2
+80q_{1234}^2q_{1567}^2 -35q^2_{1234}q^2_{5678} \rangle\ .  
$$
In a
similar way we can obtain all the other relations, from the higher
orders, simply by equating to zero all the coefficients of the
expansion in powers of $\theta$, with no need to expand in
$\beta^{\prime}$.

When $\Phi=q^{2}_{1\cdots s}$, notice that
the relation we obtain
from the lowest order in (\ref{stream}) is
formally identical to equation (\ref{generale2}),
with zero on the
left hand side, without the limit $\alpha^{\prime}\to\infty$.
Hence, using the invariance with respect to
permutations of replicas, we have the general form of
the constraint of which
(\ref{primaac}) and (\ref{secondaac}) are two special cases.
In general, for a suitable function $\Phi_{1\cdots s}$ of $s$ replicas,
from Proposition \ref{derivata} we can state
\begin{theorem} Given an even integer $s$, the AC relation
  $$
  \langle \Phi_{1\cdots s}q^{2}_{1,s+1}\rangle=
  \frac{s-1}{2s}\langle \Phi_{1\cdots s}q^{2}_{1,2}\rangle+
  \frac{s+1}{2s}\langle \Phi_{1\cdots s}q^{2}_{s+1,s+2}\rangle
  $$
  holds.
\end{theorem}
Subtracting the equation above from the well known Ghirlanda-Guerra
identity \cite{gg}
$$
\langle \Phi_{1\cdots s}q^{2}_{1,s+1}\rangle=
\frac1s\langle \Phi_{1\cdots s}\rangle\langle q^{2}_{12}\rangle
+\frac{s-1}{s}\langle \Phi_{1\cdots s}q^{2}_{s+1,s+2}\rangle
$$
we get the other well known relation
\begin{equation}\label{terzagg}
\langle \Phi_{1\cdots s}q^{2}_{s+1,s+2}\rangle=
\frac{2}{s+1}\langle \Phi_{1\cdots s}\rangle\langle q^{2}_{12}\rangle
+\frac{s-1}{s+1}\langle \Phi_{1\cdots s}q^{2}_{12}\rangle\ .
\end{equation}
While the Ghirlanda-Guerra identities are a consequence of
self-averaging, in the
thermodynamic limit, of the energy density $H_{N}/N\equiv K$
$$
\langle K\Phi_{s}\rangle=\langle K\rangle
\langle \Phi_{s} \rangle\ ,
$$
the AC relations are a consequence
of stochastic stability; but they can also
be deduced from a self-averaging
relation:
$$
\mathbb{E}\Omega(K\Phi_{s})=
\mathbb{E}\Omega(K)\Omega(\Phi_{s})\ .
$$
Clearly the third relation (\ref{terzagg}) is hence a consequence of
$$
\mathbb{E}\Omega(K)\Omega(\Phi_{s})=
\langle K\rangle
\langle \Phi_{s} \rangle\ .
$$
We stress that not only the energy density can be used
to get the various relations, but several other quantities
(as long as self-averaging is preserved) would do as well.

If we consider 4-overlaps, Proposition \ref{derivata} gives
\begin{multline*}
  \frac{(s-1)(s-2)(s-3)}{4!}\langle \Phi_{1\cdots
    s}q_{1,2,3,4}^2\rangle
  -\frac{s(s-1)(s-2)}{3!} \langle 
\Phi_{1\cdots s}q_{1,2,3,s+1}^2\rangle +\\
  \frac{(s-1)s(s+1)}{4}\langle \Phi_{1\cdots
    s}q^2_{1,2,s+1,s+2}\rangle -
  \frac{s(s+1)(s+2)}{3!}\langle 
\Phi_{1\cdots s}q_{1,s+1,s+2,s+3}^2\rangle\\
  +\frac{(s+1)(s+2)(s+3)}{4!}\langle\Phi_{1\cdots
    s}q^2_{s+1,s+2,s+3,s+4}\rangle=0
\end{multline*}
which again can be deduced from a self-averaging relation too
\cite{flt,ldsf} and
should be compared with the generalization of 
Ghirlanda-Guerra relations
\begin{multline*}
  \frac{s(s-1)(s-2)(s-3)}{3!}
  \langle q_{1,2,3,4}\Phi\rangle
  -\frac{s(s-1)(s-2)}{2}\langle q_{1,2,3,s+1}\Phi\rangle \\
  +\frac{(s-1)s(s+1)}{2!}
  \langle q_{1,2,s+1,s+2}\Phi\rangle
  -\frac{s(s+1)(s+2)}{3!}
  \langle q_{1,s+1,s+2,s+3}\Phi\rangle\\
  +\langle q_{1234}\rangle\langle \Phi \rangle
  =0\ ,
\end{multline*}
which has been found in \cite{ldsf}, as a consequence of the
self-averaging of the energy density.

We do not write explicitly the general form of the constraints
deduced from Proposition \ref{derivata}, as it is a very
simple but tedious computation, which shows that the
relations are in agreement with \cite{flt}.


\subsection{Revisiting the positivity of multi-overlaps}

We here hint at how to gain a better control
of the phase transition discussed in section \ref{positivi},
using the expansion of the cavity streaming equation,
to justify from a different perspective
what is proven in \cite{gt1}: the fluctuations of the multi-overlaps
diverge at lower temperatures as the number of replicas
increases. This is a first step in the calculation
of the critical exponents of the critical behavior
of the multi-overlaps.
We only sketch the arguments, which proceed
along the lines described in \cite{abds}.

We are going to prove that the first contribution to
the average of the 2-overlap in its
$\tanh(\beta^{\prime})=\theta$ expansion is
of order two, while it is of order four for the 4-overlap, and
so on for higher order multi-overlaps, as
intuitively expected.

Let us write the streaming equation for $\langle
q_{12}\rangle_t$, with $\beta^{\prime}=\beta$,
$\alpha^{\prime}=\alpha$
$$
\partial_t \langle q_{12} \rangle_t= \alpha 
\theta^2 \langle q_{12}^2 -4
q_{12}q_{23} + 3q_{12}q_{34}\rangle_t +O(\theta^{4})\ .
$$
But $\langle
q_{12}^2\rangle_t = \langle q_{12}^2 \rangle$ because
$q^{2}_{12}$ is a filled
monomial, and it can be integrated
offering
$$
\langle q_{12} \rangle_t = \alpha \theta^2 \langle q_{12}^2
\rangle t + \alpha \theta^2 \int_0^1 dt (-4\langle 
q_{12}q_{23} \rangle_t+
3 \langle q_{12}q_{34}\rangle_t)+O(\theta^{4})\ .
$$
We now prove that the terms
inside the integral are of higher order in $\theta$.
It is enough to notice that such terms are fillable 
but not filled, so we can
expand them using the streaming equation to evaluate the leading order
at which they contribute, which is $\theta^3$ as can be deduced
from the expansions given in Appendix \ref{cavities}.
The same approach can be used for the 4-overlap; in fact
we can write
\begin{multline*}
  \partial_t \langle q_{1234}\rangle_t = \alpha \theta^2 \langle
  10q_{1234}q_{56}-16q_{1234}q_{15}+6q_{1234}q_{12}\rangle_t +\\
  \alpha\theta^4 \langle q_{1234}^2
  -16q_{1234}q_{1235}+60q_{1234}q_{1256}-
  80q_{1234}q_{1567}-35q_{1234}q_{5678}\rangle_t+O(\theta^{6})
\end{multline*}
It can be readily seen that the only contribution at the fourth
order is due to $\langle q_{1234}^2 \rangle$, given that this is
the only filled
monomial. The calculations in Appendix \ref{cavities}
show that the three
contributions from the second order in $\theta$ (i.e.
$q_{1234}q_{56}, \ q_{1234}q_{15}, \ q_{1234}q_{12}$)
contribute at orders higher than four, so the first term in the
four-replica multi-overlap expansion is positive (being a square),
in agreement with what we showed in section \ref{positivi}.
%
We notice that, while the first contribution to the
2-overlap is of order two in $\theta$, the first
contribution to the 4-overlap is of order four,
and this result extends analogously to the higher order
multi-overlaps. So it is not surprising that at the point where the
2-overlap fluctuations start diverging, 
the fluctuations of the 4-overlap
do not, and so on for higher orders, in agreement with
what is proven in \cite{gt1}.


\section{Conclusion and outlook}

We have proven the validity of Aizenman-Contucci relations for
dilute spin glasses and exhibited further relations for
multi-overlaps. Some more general relations can be found with the
same stochastic stability methods for internal energy, but also
from the cavity part of the RaMOSt trial function, by means of a
control of the response of the average of generic observables with
respect to the change of a perturbing parameter. We also showed
that the multi-overlaps undergo the same transition the 2-overlap
exhibits in fully connected models, i.e. they remain strictly
positive, below the critical temperature, if we apply an external
field and then remove it in the thermodynamic limit. The external
field is properly modulated, in diluted systems, by the degree of
connectivity of the perturbation.

The further natural development is the
study of the extension to multi-overlaps of the
self-averaging identities (known as Ghirlanda-Guerra in
fully connected models) to prove that even multi-overlaps
are non-negative with probability one in dilute spin glasses
(the identities will be derived in \cite{ldsf}).
This would extend to odd-spin interactions the replica bounds
so far rigorously valid only for even interactions, and
such a result would be important for the application
of dilute spin glasses to optimization problems like the
K-SAT.

Another development of the current work is the calculation of the
critical exponents of the multi-overlaps, which has been gained
for fully connected models in \cite{abds} with the same
techniques here shown to be fruitful in dilute models too.
We will report on critical exponents in dilute spin glasses
soon, in \cite{bdsfg}.


\appendix

\section{Formulas for the RaMOSt expansions}
\label{formulas}

Let us report for convenience the well known expansion
$$
\tanh(x)\sum\frac{2^{2n}}{2n!}(2^{2n}-1)B_{2n}x^{2n-1}
$$
where $B_{n}$ and the Bernoulli numbers, defined by
$$
\frac{x}{e^{x}-1}=\sum\frac{B_{n}x^{n}}{n!}
$$
so that
$$
\tanh(\beta)=\beta-\frac13\beta^{3}+\frac{2}{15}\beta^{5}
-\frac{17}{315}\beta^{7}+\frac{62}{2835}\beta^{9}\cdots\ .
$$

We also report here the result the computation below,
that we used in the expansions of the previous sections.

Order two:
\begin{eqnarray*}
&&\mathbb{E}\Omega(\hat{H}^{2})=\alpha^{\prime}  \\
&&\mathbb{E}\Omega^{2}(\hat{H})=
\alpha^{\prime} \langle q^{2}_{12}\rangle
\end{eqnarray*}

Order four:
\begin{eqnarray*}
&&\mathbb{E}\Omega(\hat{H}^{4})=
\alpha^{\prime} +\alpha^{\prime 2}3\\
&&\mathbb{E}\Omega(\hat{H}^{3})\Omega(\hat{H})
=\alpha^{\prime} \langle q^{2}_{12}\rangle
+\alpha^{\prime 2}3\langle q^{2}_{12}\rangle\\
&&\mathbb{E}\Omega^{2}(\hat{H}^{2})=
\alpha^{\prime} +\alpha^{\prime 2}
(1+2\langle q^{4}_{12} \rangle)\\
&&\mathbb{E}\Omega(\hat{H}^{2})\Omega^{2}(\hat{H})
=\alpha^{\prime}  \langle q^{2}_{12}\rangle+
\alpha^{\prime 2}(\langle q^{2}_{12}\rangle
+2\langle q^{2}_{12}q^{2}_{13}\rangle)\\
&&\mathbb{E}\Omega^{4}(\hat{H})
=\alpha^{\prime} \langle q^{2}_{1234}\rangle+\alpha^{\prime 2}
3\langle q^{2}_{12}q^{2}_{34}\rangle
\end{eqnarray*}

Order six:
\begin{eqnarray*}
&&\mathbb{E}\Omega(\hat{H}^{6})=\alpha^{\prime}
+\alpha^{\prime 2}15+\alpha^{\prime 3}15\\
&&\mathbb{E}\Omega(\hat{H}^{5})\Omega(\hat{H})
=\alpha^{\prime} \langle q^{2}_{12}\rangle
+\alpha^{\prime 2}15\langle q^{2}_{12}\rangle
+\alpha^{\prime 3}15\langle q^{2}_{12}\rangle\\
&&\mathbb{E}\Omega(\hat{H}^{4})\Omega(\hat{H}^{2})
=\alpha^{\prime}
+\alpha^{\prime 2}(7+8\langle q^{4}_{12}\rangle)
+\alpha^{\prime 3}(3+12\langle q^{4}_{12}\rangle)\\
&&\mathbb{E}\Omega^{2}(\hat{H}^{3})
=\alpha^{\prime} \langle q^{2}_{12}\rangle
+\alpha^{\prime 2}15\langle q^{2}_{12}\rangle
+\alpha^{\prime 3}(9\langle q^{2}_{12}\rangle
+6\langle q^{6}_{12}\rangle)\\
&&\mathbb{E}\Omega(\hat{H})\Omega(\hat{H}^{2})
\Omega(\hat{H}^{3})=
\alpha^{\prime} \langle q^{2}_{12}\rangle
+\alpha^{\prime 2}(7\langle q^{2}_{12}\rangle
+8\langle q^{2}_{12}q^{2}_{23}\rangle)\\
&&\hspace{3.5cm}+\alpha^{\prime 3}(3\langle q^{2}_{12}\rangle
+6\langle q^{2}_{12}q^{2}_{13}\rangle
+6\langle q^{2}_{12}q^{4}_{13}\rangle)\\
&&\mathbb{E}\Omega^{3}(\hat{H}^{2})=\alpha^{\prime}
+\alpha^{\prime 2}(3+12\langle q^{4}_{12}\rangle)
+\alpha^{\prime 3}(1+6\langle q^{4}_{12}\rangle
+8\langle q^{2}_{12}q^{2}_{23}q^{2}_{31}\rangle)\\
&&\mathbb{E}\Omega(\hat{H}^{4})\Omega^{2}(\hat{H})
=\alpha^{\prime} \langle q^{2}_{12}\rangle
+\alpha^{\prime 2}(7\langle q^{2}_{12}\rangle
+8\langle q^{2}_{12}q^{2}_{23}\rangle)\\
&&\hspace{3.5cm}+\alpha^{\prime 3}(3\langle q^{2}_{12}\rangle
+12\langle q^{2}_{12}q^{2}_{23}\rangle)\\
&&\mathbb{E}\Omega^{3}(\hat{H})\Omega(\hat{H}^{3})
=\alpha^{\prime} \langle q^{2}_{1234}\rangle
+\alpha^{\prime 2}(3\langle q^{2}_{1234}\rangle
+12\langle q^{2}_{12}q^{2}_{34}\rangle)\\
&&\hspace{3.5cm}+\alpha^{\prime 3}(9
\langle q^{2}_{12}q^{2}_{34}\rangle
+6\langle q^{2}_{12}q^{2}_{23}q^{2}_{34}\rangle)\\
&&\mathbb{E}\Omega^{2}(\hat{H})\Omega^{2}(\hat{H}^{2})
=\alpha^{\prime} \langle q^{2}_{12}\rangle
+\alpha^{\prime 2}(3\langle q^{2}_{12}\rangle
+8\langle q^{2}_{12}q^{2}_{23}\rangle
+4\langle q^{2}_{12}q^{2}_{1234}\rangle)\\
&&\hspace{3.5cm}+\alpha^{\prime 3}(\langle q^{2}_{12}\rangle
+2\langle q^{2}_{12}q^{4}_{34}\rangle
+4\langle q^{2}_{12}q^{2}_{23}\rangle
+8\langle q^{2}_{12}q^{2}_{23}q^{2}_{34}\rangle)\\
&&\mathbb{E}\Omega(\hat{H}^{2})\Omega^{4}(\hat{H})
=\alpha^{\prime} \langle q^{2}_{1234}\rangle
+\alpha^{\prime 2}(\langle q^{2}_{1234}\rangle
+6\langle q^{2}_{12}q^{2}_{34}\rangle
+8\langle q^{2}_{12}q^{2}_{2345}\rangle)\\
&&\hspace{3.5cm}+\alpha^{\prime 3}
(3\langle q^{2}_{12}q^{2}_{34}\rangle
+12\langle q^{2}_{12}q^{2}_{23}q^{2}_{45}\rangle)\\
&&\mathbb{E}\Omega^{6}(\hat{H})
=\alpha^{\prime} \langle q^{2}_{123456}\rangle
+\alpha^{\prime 2}15\langle q^{2}_{12}q^{2}_{3456}\rangle
+\alpha^{\prime 3}15\langle q^{2}_{12}q^{2}_{34}q^{2}_{56}\rangle\\
\end{eqnarray*}


\section{Formulas for the cavity expansions}
\label{cavities}

In this appendix the streaming equations for the expansion of
section \ref{barra} follow:

\begin{eqnarray*}
&&
\partial_t\langle q_{12}\rangle_t= \alpha \theta^2 \langle q_{12}^2
-4q_{12}q_{23}+3q_{12}q_{34}\rangle_t+O(\theta^{4})\\
&&
\partial_t \langle q_{12}q_{23} \rangle_t = \alpha \theta^2 \langle 6
q_{12}q_{23}q_{45}-6q_{12}q_{23}q_{14}-
3q_{12}q_{23}q_{24}+q_{12}q_{23}q_{13}\\
&&
 +2q_{12}^2q_{23} \rangle  + \alpha \theta^4 \langle
15q_{12}q_{23}q_{4567}-20q_{12}q_{23}
q_{1456}-10q_{12}q_{23}q_{2456}\\
&&
 +12q_{12}q_{23}q_{1245}
+6q_{12}q_{23}q_{1345}-3q_{12}q_{23}q_{1234}\rangle_t+O(\theta^{6})
\end{eqnarray*}
\begin{eqnarray*}
&&
\partial_t \langle q_{12}q_{34} \rangle_t = \alpha \theta^2 \langle 10
q_{12}q_{34}q_{56}-16q_{12}q_{34}
q_{15}+2q_{12}^2q_{34}+4q_{12}q_{34}q_{13}\rangle_t\\
&&
+\alpha \theta^4 \langle 35 q_{12}q_{34}q_{5678}-80
q_{12}q_{34}q_{1567}+20 q_{12}q_{34}q_{1256}+40
q_{12}q_{34}q_{1356}\\
&&
-16q_{12}q_{34}q_{1235}+q_{12}q_{34}q_{1234}\rangle_t+O(\theta^{6})
\end{eqnarray*}
\begin{eqnarray*}
&&
\partial_t \langle q_{1234}\rangle_t = \alpha \theta^2 \langle
10q_{1234}q_{56}-16q_{1234}q_{15}+6q_{1234}q_{12}\rangle_t +\alpha
\theta^4 \langle q_{1234}^2\\
&&
-16q_{1234}q_{1235}+60q_{1234}q_{1256}-
80q_{1234}q_{1567}-35q_{1234}q_{5678}\rangle_t+O(\theta^{6})
\end{eqnarray*}
\begin{eqnarray*}
&&
\partial_t\langle q_{1234}q_{12} \rangle_t = \alpha \theta^2 \langle 10
q_{1234}q_{12}q_{56}-8q_{1234}q_{12}q_{15}\\
&&
-8q_{1234}q_{12}q_{35}
+q_{1234}q_{12}^2+5q_{1234}q_{12}q_{13}\rangle_t + \alpha \theta^4
\langle 35q_{1234}q_{12}q_{5678}\\
&&
-40q_{1234}q_{12}q_{1567}-40q_{1234}q_{12}q_{3567}
+10q_{1234}q_{12}q_{1256}+40q_{1234}q_{12}q_{1356}\\
&&
+10q_{1234}q_{12}q_{3456}-16q_{1234}
q_{12}q_{1235}+q_{1234}^2q_{12}\rangle_t+O(\theta^{6})
\end{eqnarray*}
\begin{eqnarray*}
&&
\partial_t \langle q_{1234}q_{15} \rangle_t =
\alpha \theta^2 \langle 15
q_{1234}q_{15}q_{67}-15q_{1234}q_{15}
q_{26}-10q_{1234}q_{15}q_{56}\\
&&
+5q_{1234}q_{15}q_{12}+4q_{1234}q_{15}
q_{23}+q_{1234}q_{15}^2\rangle_t + \alpha \theta^4 \langle 70
q_{1234}q_{15}q_{6789}\\
&&
-105q_{1234}q_{15}q_{2678}
-70q_{1234}q_{15}q_{1678}+15q_{1234}q_{15}q_{1567}+90
q_{1234}q_{15}q_{1267}\\
&&
+45q_{1234}q_{15}q_{2367}+20q_{1234}q_{15}q_{1235}
-25q_{1234}q_{15}q_{1246}\\
&&
-5q_{1234}q_{15}q_{2346}+
4q_{1234}q_{15}q_{1236}+q_{1234}^2q_{15}\rangle_t+O(\theta^{6})
\end{eqnarray*}
\begin{eqnarray*}
&&
\partial_t \langle q_{1234}q_{56}\rangle_t = \alpha \theta^2 \langle
21q_{1234}q_{56}q_{78}-24q_{1234}q_{56}q_{17}\\
&&
-12q_{1234}q_{56}q_{57}+6q_{1234}q_{56}
q_{12}+8q_{1234}q_{56}q_{57}+q_{1234}q_{56}^2\rangle\\
&&
+\alpha \theta^4 \langle 126q_{1234}q_{56}q_{7890}-112q_{1234}
q_{56}q_{5789}-224q_{1234}q_{56}q_{1789}\\
&&
+21q_{1234}q_{56}q_{5678}
+168q_{1234}q_{56}q_{1578}+126q_{1234}q_{56}
q_{1378}-24q_{1234}q_{56}q_{3567}\\
&&
+14q_{1234}q_{56}q_{1235}+q_{1234}^2q_{56}\rangle_t+O(\theta^{6})
\end{eqnarray*}

\section*{Acknowledgements}

The authors are extremely grateful to Peter Sollich for
reviewing the manuscript, and to an anonymous referee
for pointing out a weak point in a previous version of this
article.


\end{document}